\documentclass[11pt]{book}

\usepackage{jfaa2e}     
\usepackage[latin1]{inputenc} 
\usepackage{amssymb}
\usepackage{amsfonts} 
\usepackage{graphicx} 
\usepackage{amsthm} 
\usepackage{amsmath,amscd} 
\usepackage{graphicx}
\usepackage{srcltx}
\usepackage{hyperref}

\def\ni{\noindent}
\def\nn{\nonumber}

\def \bc {\begin{center}}
\def \ec {\end{center}}
\def \bi {\begin{itemize}}
\def \ei {\end{itemize}}

\def \ba {\begin{array}}
\def \ea {\end{array}}

\def \bea {\begin{eqnarray}}
\def \eea {\end{eqnarray}}

\def \be {\begin{equation}}
\def \ee {\end{equation}}

\def \ca {{\cal A}}

\def \ct {{\cal T}}
\def \cb {{\cal B}}

\def \cn {{\cal N}}
\def \ch {{\cal H}}
\def \cq {{\cal Q}}

\def \bz {\bar{z}}

\newtheorem{thm}{Theorem}[section]
\newtheorem{cor}[thm]{Corollary}
\newtheorem{lem}[thm]{Lemma}
\newtheorem{prop}[thm]{Proposition}
\newtheorem{defn}[thm]{Definition}
\theoremstyle{remark}
\newtheorem{rem}[thm]{Remark}

\newcommand{\li}{\langle}
\newcommand{\ld}{\rangle}

\begin{document}

\author{M. Calixto, J. Guerrero and J.C. S\'anchez-Monreal}


\chapter{Sampling Theorem and Discrete Fourier Transform on the Hyperboloid}

\footnotetext{\textit{Math Subject Classifications.}
                    32A10, 42B05, 94A12, 94A20,  81R30.}

\footnotetext{\textit{Keywords and Phrases.}
                     Holomorphic Functions, Coherent States, Discrete Fourier Transform,
             Sampling, Discrete Frames}

\begin{abstract}
Using Coherent-State (CS) techniques, we prove a sampling theorem
for holomorphic functions on the hyperboloid (or its stereographic
projection onto the open unit disk $\mathbb D_1$), seen as a
homogeneous space of the pseudo-unitary group $SU(1,1)$. We
provide a reconstruction formula for bandlimited functions,
through a sinc-type kernel, and a discrete Fourier transform from
$N$ samples properly chosen. We also study the case of
undersampling of band-unlimited functions and the conditions under
which a partial reconstruction from $N$ samples is still possible
and the accuracy of the approximation, which tends to be exact in
the limit $N\to\infty$.
\end{abstract}

\section{Introduction}

In a previous article \cite{samplingsphere}, we proved sampling
theorems and provided discrete Fourier transforms for holomorphic
functions on the Riemann sphere, using the machinery of Spin CS
related to the special unitary group $SU(2)$, which is the double
cover of the group $SO(3)$ of motions of the sphere $\mathbb S^2$.
Here we study similar discretization problems for its noncompact
counterpart $SO(2,1)$ (the group of motions of the Lobachevsky
plane) or, more precisely, for its double cover $SU(1,1)$. Both,
$SU(2)$ and $SU(1,1)$, appear as the underlying symmetry groups of
many physical systems for which they constitute a powerful
computational and classification tool. In fact, Angular Momentum
Theory proves to be essential when studying systems exhibiting
rotational invariance (isotropy). In the same manner, the
representation theory of $SU(1,1)$ or $SL(2,\mathbb R)$ is useful
when dealing with systems bearing conformal invariance, specially
in two dimensions, where this finite-dimensional symmetry can be
promoted to an infinite-dimensional one (the Virasoro group). In
particular, the group $SL(2,\mathbb R)$ was used in \cite{acha} to
define wavelets on the circle and the real line in a unified way.
Furthermore, $SU(2)$ and $SU(1,1)$ CS, generalizing canonical CS
of the Heisenberg-Weyl group (Gabor frames), find a great variety
of applications, mainly in the study of quantum mechanical systems
and their classical limit (see e.g. \cite{CS,Klauder,Wigner}). For
example, ground states of superconductors and superfluids (like
Bose-Einstein condensates) are coherent states. Likewise, the
Lowest Landau Level (LLL) wavefunctions in Quantum Hall Effect
(characterized by a quantization of the Hall conductance in
two-dimensional electron systems subjected to low temperatures and
strong magnetic fields) are coherent states; the formulation of
such interesting effect on the hyperboloid $SU(1,1)/U(1)$ has been
recently considered in \cite{HallDisk} (see also references
therein for the extension to other geometries) and we believe that
our construction of discrete frames and sampling theorems on
$\mathbb D_1$ can be useful when considering numerical simulations
of these systems.

Standard Continuous Wavelet Theory (see e.g. \cite{Fuhr}) can also
be seen as a chapter of CS on the group of affine transformations
(translations and dilations). Here the discretization process
turns out to be essential for computational applications in, for
example, signal processing. These results revived interest in the
question of discretization and we hope that the establishment of
new sampling theorems for harmonic analysis on non-Abelian groups
and their homogeneous spaces will be of importance for numerical
study and simulation of those physical systems bearing that
symmetries. Actually, there are some important general results
about sampling and efficient computation of Fourier transforms for
compact groups (see e.g. \cite{M1,M2}). However, a comprehensive
study of the non-compact case is far more involved, although there
is a quite well developed theory of sampling on Riemannian
manifolds (see Refs. from \cite{Pesenson1} to \cite{Stenzel}) with
reconstruction formulas for bandlimited functions on homogeneous
spaces. Other results in this direction have been obtained for
specific groups (see e.g. \cite{Gazeau} for a survey). For
instance, we would like to point out Ref. \cite{Ch1} for the
motion group and its engineering applications \cite{Ch3} (namely
in robotics \cite{Ch2}) and \cite{discretePoincare} for discrete
frames of the Poincaré group and its potential applications to
Relativity Theory.

This article intends to be a further step in this direction.
Completeness criteria for CS subsystems related to discrete
subgroups of $SU(1,1)$ have been proved using the theory of
Automorphic Forms (see e.g. \cite{CS}). Here we shall follow a
different approach.  Working in the open unit disk
$\mathbb{D}_1=SU(1,1)/U(1)$ (as an appropriate realization of the
Lobachevsky plane or the hyperboloid), we shall choose as sampling
points for analytic functions inside $\mathbb{D}_1$ (carrying a
unitary irreducible representation of $SU(1,1)$ of Bargmann index
$s$) a set of $N$ equally distributed points on a circumference of
radius $r<1$. For bandlimited holomorphic functions on
$\mathbb{D}_1$ of bandlimit $M<N$ and index $s$, the resolution
operator $\ca$ is diagonal, providing a reconstruction formula by
means of a (left) pseudoinverse. The Fourier coefficients can be
obtained by means of the (filtered) Fourier transform of the data,
allowing for a straightforward fast extension of the
reconstruction algorithm. The reconstruction of arbitrary
(band-unlimited) functions is not exact for a finite number $N$ of
samples. However, for fast-decaying, or ``quasi-bandlimited",
functions it is still possible to give partial reconstruction
formulas and to analyze the accuracy of the approximation in terms
of $N$, the radius $r$ and the index $s$, this time through the
sampled CS overlap (or reproducing kernel) $\cb$ (see later on
Sec. \ref{CSandFrames} for definitions), which exhibits a
``circulant'' structure and can be easily inverted using the
properties of the Rectangular Fourier Matrices (RFM) and the
theory of Circulant Matrices \cite{circulante}. This helps us to
provide a reconstruction formula accomplished through an
eigen-decomposition $\cb={\cal F}D {\cal F}^{-1}$ of $\cb$, where
${\cal F}$ turns out to be the standard discrete Fourier transform
matrix.

The plan of the article is as follows. In order to keep the
article as self-contained as possible, we shall introduce in the
next section general definitions and results about CS and frames
based on a group $G$. The standard construction of CS related to
the discrete series representations of $G=SU(1,1)$ is briefly
sketched in Sec. \ref{su11rep}. We refer the reader to Refs.
\cite{CS,Klauder,Gazeau,Holschneider} for more information. In
Section \ref{sampling} we provide sampling theorems, discrete
Fourier transforms and reconstruction formulas for bandlimited
holomorphic functions on $\mathbb D_1$ of bandlimit $M$ and index
$s$. For band-unlimited functions these reconstruction formulas
are not exact for finite $N$ and we analyze the error committed in
terms of $N,r$ and $s$, which tends to zero for high values of
$N$. Finally, Sec. \ref{comments} is devoted to conclusions and
outlook.

\section{Coherent States, Frames and Discretization}
\label{CSandFrames}

Let us consider a \emph{unitary} representation $U$ of a Lie group $G$ on
a Hilbert space $({\cal H},\langle \cdot|\cdot\rangle)$. Consider also the
space
$L^2(G,dg)$ of square-integrable complex functions $\Psi$ on $G$, where
$dg=d(g'g),\,\forall g'\in G$, stands for the left-invariant
Haar measure, which defines the scalar product
\be
\left(\Psi|\Phi\right)=\int_G\bar{\Psi}(g)\Phi(g)dg. \ee A non-zero
function $\gamma\in {\cal H}$ is called \emph{admissible}  (or a
\emph{fiducial} vector) if $\Gamma(g)\equiv \langle
U(g)\gamma|\gamma\rangle\in L^2(G,dg)$, that is, if
\be
c_\gamma=\int_G\bar{\Gamma}(g)\Gamma(g)dg=\int_G|\langle
U(g)\gamma|\gamma\rangle|^2dg<\infty. \label{norm}
\ee

A unitary representation for which admissible vector exists is
called \textit{square integrable}. For a square integrable
representation, besides Eq. (\ref{norm}) the following property
holds (see \cite{GMP}):
\be
\int_G|\langle
U(g)\gamma|\psi\rangle|^2dg<\infty\,,\forall \psi\in {\cal H} \,.\label{norm2}
\ee

Let us assume that the representation $U$ is \emph{irreducible}, and that
there exists a function $\gamma$ admissible, then a system of coherent
states (CS) of ${\cal H}$  associated to (or indexed by) $G$ is defined
as the set of functions in the orbit of $\gamma$ under $G$
\be
\gamma_g=U(g)\gamma, \;\; g\in G.
\ee

There are representations without admissible vectors, since the integration with respect to some subgroup diverges.
In this case, or even for convenience when admissible vectors exist, we can restrict ourselves to a suitable
homogeneous space $Q=G/H$, for some closed subgroup $H$. Then, the
non-zero function $\gamma$ is said to be admissible mod$(H,\sigma)$ (with
$\sigma:Q\to G$ a Borel section\footnote{A section $\rho :Q\to G$
of the fibre bundle $G \stackrel{H}{\to}Q$ with base $Q$ and fibre $H$ is said to be a Borel
section if it is measurable with respect to
the Borel $\sigma$-algebras of $Q$ and $G$.}),
and the representation $U$ square
integrable mod$(H,\sigma)$, if the condition
\be
\int_Q|\langle U(\sigma(q))\gamma|\psi\rangle|^2 d q<\infty,\;\;\forall
\psi\in {\cal H}\label{qsquare}
\ee
holds, where $d q$ is a measure on $Q$
``projected'' from the left-invariant measure $dg$ on the whole $G$ (see \cite{medida}). Note that this more general
definition of square integrability includes the previous one for $H=\{e\}$ and $\sigma$ the identity function since Eq. (\ref{qsquare}) reduces to Eq. (\ref{norm2}), and this implies the square integrability condition (\ref{norm}).

The
coherent states indexed by $Q$ are defined as
$\gamma_{\sigma(q)}=U(\sigma(q))\gamma, q\in Q$, and they form an
overcomplete set in ${\cal H}$.

The condition (\ref{qsquare}) could also be written as an
``expectation value" \be 0<\int_Q |\langle
U(\sigma(q))\gamma|\psi\rangle|^2 dq=\langle \psi|A_\sigma
|\psi\rangle <\infty ,\;\;\forall \psi\in {\cal
H},\label{pbiop}\ee where
$A_\sigma=\int_Q|\gamma_{\sigma(q)}\rangle\langle
\gamma_{\sigma(q)}|dq$ is a positive, bounded, invertible
operator.\footnote{In this paper we shall extensively use the
Dirac notation in terms of ``bra'' and ``kets'' (see e.g.
\cite{acha,Gazeau}). The Dirac notation is justified by the Riesz
Representation Theorem, and is valid in more general settings than
Hilbert spaces of square integrable functions .}

If the operator $A_\sigma^{-1}$ is also bounded,
then the set $S_\sigma=\{|\gamma_{\sigma(q)}\rangle, q\in Q\}$ is called a
\emph{frame} (see \cite{frames} for details on frames), and a \emph{tight frame} if $A_\sigma$ is a positive
multiple of the identity, $A_\sigma=\lambda {I}, \lambda>0$.

To avoid domain problems in the following, let us assume that $\gamma$
generates a frame (i.e., that $A_\sigma^{-1}$ is bounded). The \emph{CS
map} is defined as the linear map \be\begin{array}{cccc} T_\gamma: & {\cal
H}&\longrightarrow&
L^2(Q,dq)\\
 & \psi & \longmapsto & \Psi_\gamma(q)=[T_\gamma\psi](q)=\frac{\langle
\gamma_{\sigma(q)}|\psi\rangle}{\sqrt{c_\gamma}}.\end{array}
\label{cwt}\ee  Its range $L^2_\gamma(Q,dq)\equiv T_\gamma({\cal
H})$ is complete with respect to the scalar product
$(\Phi|\Psi)_\gamma\equiv\left(\Phi|T_\gamma A_\sigma^{-1}
T_\gamma^{-1}\Psi\right)_Q$ and $T_\gamma$ is unitary from ${\cal
H}$ onto $L^2_\gamma(Q,dq)$. Thus, the inverse map $T_\gamma^{-1}$
yields the \emph{reconstruction formula}
\be
\psi=T_\gamma^{-1}\Psi_\gamma=\int_Q\Psi_\gamma(q)A_\sigma^{-1}\gamma_{\sigma(q)}
d q,\;\;\Psi_\gamma\in L^2_\gamma(Q,d q),\ee which expands $\psi$
in terms of CS $A_\sigma^{-1}\gamma_{\sigma(q)}$ with coefficients
$\Psi_\gamma(q)=[T_\gamma\psi](q)$. These formulas acquire a
simpler form when $A_\sigma$ is a multiple of the identity, as is
for the case considered in this article.

When it comes to numerical calculations, the integral
$A_\sigma=\int_Q|\gamma_{\sigma(q)}\rangle\langle \gamma_{\sigma(q)}|d q$
has to be discretized, which means to restrict ourself to a discrete
subset ${\cal Q}\subset Q$. The question is whether this restriction will
imply a loss of information, that is, whether the set ${\cal
S}=\{|q_k\rangle\equiv|\gamma_{\sigma(q_k)}\rangle, q_k\in{\cal Q} \}$
constitutes a discrete frame itself, with resolution operator
\be {\cal
A}=\sum_{q_k\in {\cal Q}}|q_k\rangle\langle q_k|.
\ee
The operator ${\cal
A}$ need not coincide with the original ${A}_\sigma$. In fact, a
continuous tight frame might contain discrete non-tight frames, as happens
in our case (see later on Sec. \ref{sampling}).

Let us assume that ${\cal S}$ generates a discrete frame, that is,
there are two positive constants $0<b<B<\infty$ (\emph{frame
bounds}) such that the admissibility condition \be
b\|\psi\|^2\leq\sum_{q_k\in {\cal Q}} |\langle
q_k|\psi\rangle|^2\leq B\|\psi\|^2\label{pbiop2} \ee holds
$\forall \psi\in{\cal H}$. To discuss the properties of a frame,
it is convenient to define the frame (or sampling) operator ${\cal
T}:{\cal H}\to \ell^2$ given by ${\cal T}(\psi)=\{\langle
q_k|\psi\rangle, \,q_k\in{\cal Q}\}$. Then we can write ${\cal
A}={\cal T}^*{\cal T}$, and the admissibility condition
(\ref{pbiop2}) now adopts the form \be b I\leq {\cal T}^*{\cal
T}\leq B I,\ee where $I$ denotes the identity operator in ${\cal
H}$. This implies that ${\cal A}$ is invertible. If we define the
\emph{dual frame} $\{|\tilde{q}\rangle\equiv {\cal A}^{-1}
|q\rangle\}$, one can easily prove that the expansion
(\emph{reconstruction formula})
\be |\psi\rangle=\sum_{q_k\in {\cal Q}} \Psi_k|\tilde{q}_k\rangle\label{reconstructionformula-over}\ee
where $\Psi_k\equiv \langle q_k|\psi\rangle$, converges
strongly in ${\cal H}$, that is, the expression \be{\cal
T}_l^+{\cal T}= \sum_{q_k\in {\cal Q}}|\tilde{q}_k \rangle\langle
q_k|= {{\cal T}}^*({\cal T}_l^+)^*= \sum_{{q}_k\in {\cal Q}}|q_k
\rangle\langle \tilde{q}_k |= I\label{resolucionidentidad}\ee
provides a resolution of the identity, where ${\cal
T}_l^+\equiv({\cal T}^*{\cal T})^{-1}{\cal T}^*$ is the (left)
pseudoinverse (see, for instance, \cite{pseudoinverse}) of ${\cal
T}$ (see e.g. \cite{Holschneider,Gazeau} for a proof, where they
introduce the dual frame operator $\tilde{\ct}=(\ct_l^+)^*$
instead).

It is interesting to note that the operator $P={\cal T}{\cal T}_l^+$ acting
on $\ell^2$ is an orthogonal projector onto the range
of $\ct$.


{}From (\ref{reconstructionformula-over}) the function $\Psi(q)$
can be obtained
\be \Psi(q)\equiv \langle q|\psi\rangle =
\sum_{q_k\in {\cal Q}} \Xi_k(q)\Psi_k \ee
from its samples $\Psi_k=\langle q_k|\psi\rangle$, through some
``sinc-type'' kernel
\be \Xi_k(q)=\langle q|\tilde{q}_k\rangle\label{sinctype}
\ee
fulfilling $\Xi_k(q_l)=P_{lk}$. A projector is obtained, instead
of the identity, to account for the fact that an arbitrary set of
overcomplete data $\{\Psi_k\}\in \ell^2$, can be incompatible with
$|\psi\rangle\in {\cal H}$, and therefore they are previously
projected (note that an overdetermined system of equations is
being solved).

This case will be named \emph{oversampling}, since there are more data
than unknowns, and will be discussed in Sec.
\ref{oversampling} (see also \cite{samplingsphere}). In other
contexts, when eq.  (\ref{pbiop2}) holds, the set ${\cal Q}$ is
said to be \emph{sampling} for the space ${\cal H}$ \cite{Fuhr}.

We shall be mainly interested in cases where there are not enough
points to completely reconstruct a given function $\psi$, i.e.,
\emph{undersampling}, but a partial reconstruction is still
possible. In these cases ${\cal S}$ does not generate a discrete
frame, and the resolution operator ${\cal A}$ would not be
invertible. But we can construct another operator from ${\cal T}$,
${\cal B}={\cal T}{\cal T}^*$, acting on $\ell^2$.

The matrix elements of ${\cal B}$ are
\be
{\cal B}_{kl}=\langle q_k|q_l\rangle\,,\label{overlapping}
\ee
therefore ${\cal B}$ is the discrete reproducing kernel operator,
see eq. (\ref{CSoverlap}). If the set ${\cal S}$ is linearly
independent, the operator ${\cal B}$ will be invertible and a
(right) pseudoinverse can be constructed for ${\cal T}$, ${\cal
T}_r^+\equiv {\cal T}^*({\cal T}{\cal T}^*)^{-1}$, in such a way
that $\ct \ct_r^+ = I_{\ell^2}$. As in the previous case there is
another operator, $P_{\cal S}= \ct_r^+ \ct$ acting on ${\cal H}$
which is an orthogonal projector onto the subspace ${\cal H}^{\cal
S}$ spanned by ${\cal S}$. A pseudo-dual frame can be defined as
\be
|\tilde{q}_k\rangle = \sum_{q_l\in {\cal Q}} ({\cal
B}^{-1})_{lk}|q_l\rangle \label{discrete-repker}\ee
providing a resolution of the projector $P_{\cal S}$,
\be
 {\cal T}_r^+ \ct = \sum_{q_k\in {\cal Q}}|\tilde{q}_k \rangle\langle q_k|=
\ct^* ({\cal T}_r^+)^*  = \sum_{{q}_k\in {\cal Q}}|q_k \rangle\langle \tilde{q}_k |=
P_{\cal S} \label{resolucionproyector}
\ee
Using this, a partial reconstruction (an ``alias'') $\hat{\psi}$
of $\psi$ is obtained,
\be \hat{\Psi}(q)=\langle q| \hat{\psi}\rangle=\sum_{q_k\in {\cal
Q}} L_k(q)\Psi_k,\label{partialrec} \ee
from its samples $\Psi_k=\langle q_k|\psi\rangle$, through some
 ``Lagrange-like''interpolating functions
\be L_k(q)=\langle q|\tilde{q}_k\rangle\label{interpolating}
\ee
fulfilling $L_k(q_l)=\delta_{kl}$. The alias $\hat{\psi}$ is the
orthogonal projection of $\psi$ onto the subspace ${\cal H}^{\cal
S}$, that is, $|\hat{\psi}\rangle=P_{\cal S}|\psi\rangle$. The
relative (normalized) distance from the exact $\psi$ to the
reconstructed function $\hat{\psi}$ is given by the relative error
function:
\be {\cal E}_\psi({\cal H}^{\cal S})=\frac{\|
\psi-\hat{\psi}\|}{\|\psi\|}=\sqrt{\frac{\langle\psi|I-P_{\cal
S}|\psi\rangle}{\langle \psi|\psi\rangle}}\label{errorpsi}\ee

As mentioned above, we shall denote this case by
\emph{undersampling}, since there are not enough data to fully
reconstruct $\psi$. In other contexts, a set ${\cal Q}$ is said to
be \emph{interpolating} if, for an arbitrary set of data
$\{\Psi_k\}$ there exists a $|\psi\rangle\in {\cal H}$ such that
$\langle q_k|\psi\rangle = \Psi_k$ \cite{Fuhr}. This condition is
satisfied in this case since $L_k(q_l)=\delta_{kl}$.

The two operators ${\cal A}$ and ${\cal B}$ are intertwined by the
frame operator ${\cal T}$, ${\cal T}{\cal A}={\cal B}{\cal T}$. If
${\cal T}$ were invertible, then both ${\cal A}$ and ${\cal B}$
would be invertible and ${\cal T}_r^+={\cal T}_l^+ ={\cal
T}^{-1}$. This case would correspond to \emph{critical sampling},
where both operators ${\cal A}$ and ${\cal B}$ can be used to
fully reconstruct the function $\psi$. However, in many cases it
is not possible to find a set of points $\emph{Q}$ such that both
${\cal A}$ and ${\cal B}$ are invertible, that is, there is no
critical sampling, or there are not sets $\emph{Q}$ which are
sampling and interpolating at the same time. The most common
example is the Bargmann-Fock space of analytical functions on
$\mathbb{C}$, where one can find rectangular lattices which are
sampling (and therefore ${\cal A}$ is invertible), or which are
interpolating (and thus ${\cal B}$ is invertible), but not both
simultaneously \cite{CS,Fuhr}. Examples of critical sampling are
given by the space of band limited functions on $\mathbb{R}$ and
the set $\mathbb{Z}$, which is both sampling and interpolating,
and the space of functions on the Riemann sphere (or rather its
stereographic projection onto the complex plane) with fixed
angular momentum $s$ and the set of $N^{\rm th}$-roots of unity,
with $N=2s+1$ \cite{samplingsphere}.

It should be noted that in the case in which there is a finite number $N$ of
sampling points $q_k$, the space $\ell^2$ should be substituted by
$\mathbb{C}^N$, and the operator $\cb$ can be identified with its matrix
once a basis has been chosen.

\section{Representations of $SU(1,1)$: Discrete Series\label{su11rep}}

Discrete series representations of $SU(1,1)$ can be found in the
literature \cite{CS,Klauder}. Here we shall try to summarize what is
important for our purposes, in order to keep the article as self-contained
as possible.

\subsection{Coordinate Systems and Generators}

The group $SU(1,1)$ consists of all unimodular $2\times 2$ matrices
leaving invariant the pseudo-Euclidean metric $\eta={\rm diag}(1,-1)$ and
can be parametrized as
\be
SU(1,1)=\{ U(\zeta)=\left(
\begin{array}{cc}\zeta_1&\zeta_2\\\bar{\zeta}_2&\bar{\zeta}_1\end{array}\right), \,\,
\zeta_{1},\zeta_{2} \in {\mathbb C}:
\det(U)=|\zeta_{1}|^{2}-|\zeta_{2}|^{2}=1 \}.\label{su11group} \ee
The group $SU(1,1)$ is locally isomorphic to the three-dimensional Lorentz
group $SO(2,1)$ (the group of ``rotations'' of the three-dimensional
pseudo-Euclidean space). More precisely
$SO(2,1)=SU(1,1)/\mathbb Z_2$, where $\mathbb Z_2=\{I,-I\}$ ($I$ is the $2\times 2$ identity matrix) is the cyclic group
with two elements. The group $SU(1,1)$ acts on $\mathbb C$ as
\be
U(\zeta):\mathbb C\rightarrow \mathbb C,\, z\mapsto z'=\frac{\zeta_1
z+\bar{\zeta}_2}{\zeta_2 z+\bar{\zeta_1}}. \label{su11action}\ee
This action is not transitive, so that $\mathbb C$ is foliated into three
orbits:
\be
\mathbb D_1=\{z\in\mathbb C: |z|<1\},\;\;\mathbb C-\overline{\mathbb
D}_1=\{z\in\mathbb C: |z|>1\},\;\; \mathbb S_1=\{z\in\mathbb C:
|z|=1\}.\ee
The open unit disk $\mathbb D_1$ may be considered as the stereographical
projection of the upper sheet of the two-sheet hyperboloid $\mathbb
H^2=\{(x_0,x_1,x_2)\in \mathbb R^3: x_0^2-x_1^2-x_2^2=1\}$ onto the
complex plane. The hyperboloid $\mathbb H^2$ may be identified with the
set of elements of $SU(1,1)$ with
$\zeta_1=x_0=\cosh(\tau/2)$ and
$\zeta_2=x_1+ix_2=\sinh(\tau/2)e^{i\alpha}, \tau>0, \alpha\in [0,2\pi[$,
the stereographical projection being given by
$z=\zeta_2/\zeta_1=\tanh(\tau/2)e^{i\alpha}\in \mathbb D_1$.
Thus, we could also identify $\mathbb D_1$ with the coset $SU(1,1)/U(1)$,
where $U(1)$ is the (diagonal) subgroup of the phase
$e^{i\varphi}=\zeta_1/|\zeta_1|$.

Let us consider matrices $U(\zeta)\in SU(1,1)$ in
(\ref{su11group}) near the identity $I$, i.e. $\zeta\simeq
e+\delta\zeta$ and $U(\zeta)=
I+\sum_{a}\delta\zeta_aK_a+O(\delta\zeta^2)$, where $\zeta_a,
a=0,+,-$, is a shorthand for $\zeta_1,\zeta_2,\bar{\zeta}_2$;
$e=(\zeta_0,\zeta_+,\zeta_-)=(1,0,0)$ and the infinitesimal
generators $K_a$ are:
\be  K_0=\frac{1}{2}\left(
\begin{array}{cc}
  1 & 0 \\
  0 & -1 \\
\end{array}
\right),\;\;
K_+=\left(
\begin{array}{cc}
  0 & 1 \\
  0 & 0 \\
\end{array}
\right),\;\; K_-=\left(
\begin{array}{cc}
  0 & 0 \\
  -1 & 0 \\
\end{array}
\right).\ee
They close the following Lie algebra commutation relations:
\be [K_0,K_\pm]=\pm K_\pm,\;\;[K_+,K_-]=-2K_0.\label{commurel}\ee
Although we have obtained the commutators (\ref{commurel}) from a
particular (fundamental) representation of $SU(1,1)$ in terms of
$2\times 2$ matrices, we can abstract them and look for more
general representations in terms of higher-dimensional matrices.
In particular, we are interested in a class of unitary representations of
$SU(1,1)$ which, being a non-compact group, must be infinite-dimensional. They shall be
explicitly constructed in the next section.

To finish this section, let us remind the form of the quadratic
(Casimir) operator of $SU(1,1)$:
\be C=K_0^2-(K_+K_-+K_-K_+)/2.\label{Casimir}\ee
It is not difficult to verify that $C$ commutes with every $K_a,
a=0,+,-$.

\subsection{Unitary irreducible representations: $SU(1,1)$ coherent
states} We are seeking for unitary irreducible representations of
$SU(1,1)$. By Schur's lemma, for any irreducible representation of
the Lie algebra $su(1,1)$, the Casimir operator $C$ must be a
multiple of the identity $I$, which we shall set $C=s(s-1)I$.
Thus, an irreducible representation of $SU(1,1)$ is labelled by a
single number $s$ (the Bargmann index, which we shall refer to as
the symplectic spin or just ``symplin''). We shall restrict
ourselves to discrete series representations, which are square
integrable and where $s$ is half-integer $s=1,3/2,2,5/2,\dots$. We
shall take the orthonormal basis vectors $|s,n\rangle$ in the
carrier (Hilbert) space ${\cal H}_s$ to be eigenvectors of $K_0$:
\be
K_0|s,n\rangle\equiv (n+s)|s,n\rangle.\label{ko} \ee
{}From the commutation relations (\ref{commurel}), we observe that $K_\pm$
act as raising and lowering ladder operators, respectively, whose action
on the basis vectors proves to be
\be K_+|s,n\rangle=\sqrt{(n+1)(2s+ n)}|s,n+1\rangle,\;\;
K_-|s,n\rangle=\sqrt{n(2s+n-1)}|s,n-1\rangle. \label{kpm}\ee
Indeed, it can be easily checked that the action (\ref{ko},\ref{kpm})
preserves the commutation relations (\ref{commurel}); for example:
\be
[K_+,K_-]|s,n\ld=(n(2s+n-1)-(n+1)(2s+n))|s,n\ld=-2(n+s)|s,n\ld=-2K_0|s,n\ld,\ee
and so on. From the expression (\ref{kpm}) we deduce that the spectrum of
$K_0$ is unbounded from above, that is, the Hilbert space
${\cal H}_s$ is infinite-dimensional.

Any group element $U(\zeta)\in SU(1,1)$ can also be written through the
exponential map
\be
U(z,\bar{z},\varphi)= e^{\xi K_+-\bar{\xi} K_-}e^{i\varphi
K_0},\;\;\xi=|\xi|e^{i\beta},\, z=\tanh|\xi|
e^{i\beta}.\label{Hopfcomplex} \ee
Note that the subgroup $U(1)\subset SU(1,1)$, generated by $K_0$,
stabilizes any basis vector up to an overall multiplicative phase
factor (a character of $U(1)$), i.e., $e^{i\varphi
K_0}|s,m\ld=e^{i(m+s)\varphi}|s,m\ld$. Thus, according to the
general prescription explained in Sec. \ref{CSandFrames}, letting
$Q=SU(1,1)/U(1)=\mathbb D_1$ and taking the Borel section
$\sigma:Q\to G$ with $\sigma(z,\bz)=(z,\bz,0)$, we shall define,
from now on, families of covariant coherent states ${\rm
mod}(U(1),\sigma)$ (see \cite{Gazeau}). In simple words, we shall
set $\varphi=0$ and drop it from the vectors
$U(z,\bar{z},\varphi)|s,m\ld$.

For any choice of fiducial vector $|\gamma\ld=|s,m\ld$ the set of coherent
states $|z,m\ld\equiv U(z,\bz)|\gamma\ld$ is overcomplete (for any
$m$) in ${\cal H}_s$. We shall use $|\gamma\ld=|s,0\ld$ as fiducial vector (i.e., the
lowest weight vector), so that $K_-|\gamma\ld=0$ and the coherent states
\be |z\ld\equiv U(z,\bz)|\gamma\ld=e^{\xi K_+-\bar{\xi} K_-}
|s,0\ld=\cn_s(z,\bar{z}) e^{zK_+}|s,0\ld,\ee
are holomorphic (only a function of
$z$), apart from the normalization factor
$\cn_s$ which can be determined as follows.
Exponentiating the relations (\ref{kpm}) gives
\bea e^{zK_+}|s,0\ld &=& |s,0\ld
+z\sqrt{2s}|s,1\ld+\frac{1}{2}z^2\sqrt{2s}\sqrt{2(2s+1)} |s,2\ld
+\dots\nn\\ &=& \sum_{n=0}^\infty \binom{2s+n-1}{n}^{1/2}z^n|s,n\rangle
\equiv\cn_s(z,\bz)^{-1}|z\ld.\label{exp-ansion}\eea Then, imposing
unitarity, i.e., $\li z|z\ld=1$, we arrive at
$\cn_s(z,\bz)=(1-|z|^2)^s$.

The frame $\{|z\ld, z\in\mathbb C\}$ is tight in
${\cal H}_s$, with resolution of unity
\be I=\frac{2s-1}{\pi}\int_{\mathbb D_1}|z\ld \li
z|\frac{d^2z}{(1-z\bz)^2},\label{resolholo}\ee where we denote
$d^2z=d{\rm Re}(z)d{\rm Im}(z)$. Indeed, using (\ref{exp-ansion}) we have
that
\bea \frac{2s-1}{\pi}\int_{\mathbb D_1}|z\ld \li
z|\frac{d^2z}{(1-z\bz)^2}&=& \frac{2s-1}{\pi}\int_{\mathbb D_1}
\cn_s(z,\bz)^2 \sum_{n,m=0}^{\infty}\sqrt{
\binom{2s+n-1}{n}\binom{2s+m-1}{m}}\nn  \\ & & z^n\bz^m |s,n\ld\li
s,m| \frac{ d{\rm Re}(z)d{\rm Im}(z)}{(1-z\bz)^{2}} \nn \\ &=&
2(2s-1)\sum_{n=0}^{\infty} \binom{2s+n-1}{n}|s,n\ld\li s,n|
\int_0^1 (1-r^2)^{2s-2} r^{2n+1} dr \nn \\ &=&
\sum_{n=0}^\infty|s,n\ld\li s,n|=I, \eea
where polar coordinates were used at intermediate stage.

Using (\ref{exp-ansion}), the decomposition of the coherent state $|z\ld$
over the orthonormal basis $\{|s,m\ld\}$ gives the irreducible matrix
coefficients
\bea \li z|s,m\ld &=&\li
s,0|U(z,\bz)^*|s,m\ld=\tbinom{2s+m-1}{m}^{1/2}(1-z\bz)^{s}\bz^{m}\nn
\\ &\equiv& U^s_m(z).\label{upsilon}\eea
A general symplin $s$ function
\be |\psi\ld= \sum_{m=0}^{\infty}a_m |s,m\ld\ee
is represented in the present complex characterization by
\be {\Psi}(z)\equiv\li z|\psi\ld=\sum_{m=0}^{\infty}a_m
U^s_m(z),\ee which is an anti-holomorphic function of
$z$.\footnote{Here we abuse notation when representing the
non-analytic functions $U^s_m$ and $\Psi$ simply as $U^s_m(z)$ and
$\Psi(z)$, which are indeed (anti-)holomorphic up to the
normalizing, non-analytic (real), pre-factor $\cn_s=(1-z\bz)^{s}$.
Usually, this pre-factor is absorbed into the integration measure
in (\ref{resolholo}). } The Fourier coefficients $a_n$ can be
calculated through the following integral formula:
\be a_n=\li s,n|\psi\ld=\frac{2s-1}{\pi}\int_{\mathbb
D_1}\Psi(z)\bar{U}^s_n(z)\frac{d^2z}{(1-z\bz)^2}.\label{CFT}\ee

Note that the set of CS $\{|z\ld\}$ is not orthogonal. The CS overlap (or
Reproducing Kernel) turns out to be
\be C(z,z')=\langle
z|z'\rangle=\frac{(1-z\bar{z})^s(1-z'\bar{z}')^s}{(1-z'\bar{z})^{2s}}.\label{CSoverlap}
\ee
This quantity will be essential in our sampling procedure on the disk
$\mathbb D_1$.

There are other pictures of CS for $SU(1,1)$ corresponding to other
parameterizations like, for example, the one that takes $\mathbb D_1$ to
the upper complex plane, but we shall not discuss them here.

\section{Sampling Theorem and DFT on $\mathbb D_1$\label{sampling}}

Sampling techniques consist in the evaluation of a continuous
function (a ``signal'') $\psi$ on a discrete set of points and
later (fully or partially) recovering $\psi$ without losing
essential information in the process, and the criteria to that
effect are given by various forms of Sampling Theorems. Basically,
the density of sampling points must be high enough to ensure the
reconstruction of the function in arbitrary points with reasonable
accuracy. We shall concentrate ourselves on symplin-$s$
holomorphic functions and sample them at appropriate points.

In our case, there is a convenient way to select the sampling
points in such a way that
 the resolution operator ${\cal A}$ and/or the reproducing kernel operator ${\cal B}$ are invertible
 and  explicit formulas for their inverses
are available. These are given by the set of $N$ points uniformly
distributed on a circumference of radius $r$:
\be{\cal Q}=\{q_k=z_k=r e^{2\pi i k/N}, k=0,1,\dots N-1\}\,,
r\in(0,1)\label{discreteQ}
\ee
which is a discrete subset of the homogeneous space
$Q=SU(1,1)/U(1)=\mathbb D_1$, made of the $N^{\rm th}$ roots of
$r^N$, with $0<r<1$\footnote{In the case of the sphere \cite{samplingsphere} the sampling points were
the $N^{\rm th}$-roots of unity, which formed an abelian subgroup $\mathbb{Z}_N$ of $SU(2)$,
and the main advantage of this was that  $\cb$ was a circulant matrix. In that case the sampling was
\emph{regular} \cite{Fuhr}. Now the sampling  is \emph{irregular} since $\cq$ is not a subgroup,
although it is the orbit of the subgroup
$\mathbb{Z}_N$ through the point $z=r$, and this is enough to keep the circulant structure of ${\cal B}$.
}. Denote by ${\cal S}=\{
|z_k\rangle\,,k=0,1,\ldots,N-1\}$ the subset of coherent states
associated with the points in ${\cal Q}$ and by
\be {\cal H}_s^{\cal S}\equiv  {\rm
Span}(|z_0\rangle,|z_1\rangle,\dots,|z_{N-1}\rangle)\label{HsS}
\ee
the subspace of ${\cal H}_s$ spanned by ${\cal S}$. For finite $N$
we have ${\cal H}_s^{\cal S}\not={\cal H}_s$, so that we cannot
exactly reconstruct every function $\psi\in {\cal H}_s$ from $N$
of its samples $\Psi_k=\langle z_k|\psi\rangle$, but we shall
prove that for \emph{bandlimited functions}
 we can always provide an exact
reconstruction formula.

\subsection{Bandlimited Functions\label{oversampling}}

{\defn \label{bandlimited} We define the subspace ${\cal H}_s^M$
of bandlimited functions of bandlimit $M<\infty$ as:
\be  {\cal H}_s^M\equiv {\rm
Span}(|s,0\rangle,|s,1\rangle,\dots,|s,M\rangle).\label{band-lim-space}\ee
} The subspace ${\cal H}_s^M$ is clearly a finite dimensional
vector subspace of ${\cal H}_s$, but it is not invariant under the
action of $SU(1,1)$. It is, however, invariant under the action of
the subgroup $U(1)\subset SU(1,1)$ generated by $K_0$ and, in this
sense, it resembles the space of bandlimited functions on
$\mathbb{R}$, which is invariant under the Abelian group
$\mathbb{R}$.

{\thm \label{maintheorem-over} Given a bandlimited function
$\psi\in {\cal H}_s^M$ on the disk $\mathbb D_1$, of band limit
$M$, with a finite expansion
\be |\psi\rangle=\sum_{m=0}^M a_m|s,m\rangle,\label{FourExpan}\ee
there exists a reconstruction formula
(\ref{reconstructionformula-over}) of $\psi$
\be {\Psi}(z)=\sum_{k=0}^{N-1}
\Xi_k(z)\Psi_k,\label{reconstruccionover} \ee
from $N>M$ of its samples $\Psi_k$ taken at the sampling points
in (\ref{discreteQ}), through a ``sinc-type'' kernel  given by
\be
\Xi_k(z)=\frac{1}{N}\left(\frac{1-z\bar{z}}{1-z_k\bar{z}_k}\right)^s\sum_{m=0}^M(\overline{z
z_k^{-1}})^{m}.\label{xiover}\ee }
Firstly, we shall introduce some notation and prove some previous
lemmas.

{\lem \label{lemmaover1}  For $N>M$, the frame operator $\ct:{\cal
H}_s^M\to \mathbb C^N$ given by $\ct(\psi)=\{\langle
z_k|\psi\rangle, z_k\in {\cal Q}\}$ is such that the resolution
operator ${\cal A}=\ct^*\ct$ is diagonal, ${\cal A}={\rm
diag}(\lambda_0,\ldots,\lambda_{M})$, in the basis
(\ref{band-lim-space}) of ${\cal H}_s^M$, with\footnote{The
quantities $\lambda_m$ are well defined for $m\in \mathbb{N}\cup
\{0\}$ and they will be used in the case of band-unlimited
functions.}
\be \lambda_m\equiv
N(1-r^2)^{2s}\binom{2s+m-1}{m}r^{2m},\,m=0,\dots,M.\label{lambdaover}\ee
Hence, ${\cal A}$ is invertible in ${\cal H}_s^M$. Therefore,
denoting $|\tilde{z}_k\rangle\equiv {\cal A}^{-1}|z_k\rangle$, the
dual frame, the expression
\be I_{M}=\sum_{k=0}^{N-1}|{z}_k\rangle \langle \tilde{z}_k| =
\sum_{k=0}^{N-1}|\tilde{z}_k\rangle \langle {z}_k |
\label{resolident}\ee
provides a resolution of the identity in ${\cal H}_s^M$.
\label{lemaover}}

\ni \textbf{Proof.} Using (\ref{upsilon}), the matrix elements of
${\cal T}$ can be written as:
\be {\cal T}_{kn}=\langle
z_k|s,n\rangle=\binom{2s+n-1}{n}^{1/2}(1-r^2)^sr^ne^{-2\pi
ikn/N}=\lambda_n^{1/2}{\cal F}_{kn},\label{calt}\ee
where ${\cal F}$ denotes the Rectangular Fourier Matrix (see
\cite{samplingsphere}) given by ${\cal
F}_{kn}=\frac{1}{\sqrt{N}}e^{-i2\pi kn/N}, k=0,\dots,N-1,
n=0,\dots,M$.\footnote{For the sake of briefness, we shall use
here the same notation for Rectangular Fourier Matrices as for the
square ones, namely ${\cal F}$, in the hope that no confusion
arises (see Appendix A of \cite{samplingsphere} for a more precise
distinction between both cases).} Then, the matrix elements of the
resolution operator turn out to be\footnote{Here and in the
following we shall abuse notation denoting by $A_{ij}^*$ the
element $(A^*)_{ij}=\bar{A}_{ji}$ of the hermitian conjugate
matrix $A^*$ of $A$.}
\be {\cal A}_{nm}=\sum_{k=0}^{N-1} {\cal T}_{nk}^*{\cal T}_{km}
=\lambda_n^{1/2}\lambda_m^{1/2}\sum_{k=0}^{N-1}{\cal
F}^*_{nk}{\cal F}_{km}=\lambda_n\delta_{nm}\,, \ee
where we have used the well known orthogonality relation for
Rectangular Fourier Matrices (RFM) (see e.g. Appendix A of
\cite{samplingsphere} for a discussion of some of their
properties):
\bea N\sum_{k=0}^{N-1}{\cal F}^*_{nk}{\cal
F}_{km}&=&\sum_{k=0}^{N-1}\left( e^{2\pi
i(n-m)/N}\right)^k=\left\{\ba{l} N,\;{\rm if}\; (n-m) \,{\rm
mod}\, N = 0\, \\ 0,\;{\rm if}\; (n-m)\,{\rm mod} \,N \not=0
\,\,\ea\right\}\nn\\ &=&N \delta_{(n-m) \,{\rm mod}
N,0},\label{exponencial}\eea
and this equals $N\delta_{n,m}$ if $N>M$. Since ${\cal A}$ is
diagonal with
non-zero diagonal elements $\lambda_n$, then it is invertible and
a dual frame and a (left) pseudoinverse for $\ct$ can be
constructed, ${\cal T}_l^+\equiv \ca^{-1}{\cal T}^*$, providing,
according to eq. (\ref{resolucionidentidad}), a resolution of the
identity. $\blacksquare$
\\
\ni \textbf{Proof of Theorem \ref{maintheorem-over}.} From the
resolution of the identity (\ref{resolident}), any $\psi\in {\cal
H}_s^M$ can be written as $|\psi\rangle = \sum_{k=0}^{N-1}\Psi_k
|\tilde{z}_k\rangle$, and therefore $\Psi(z)=\langle
z|\psi\rangle= \sum_{k=0}^{N-1}\Psi_k\langle
z|\tilde{z}_k\rangle$. Using that $|\tilde{z}_k\rangle =
\ca^{-1}|z_k\rangle$, we derive that \bea \langle
z|\tilde{z}_k\rangle &=& \sum_{m=0}^M \langle z|s,m\rangle
(\ca^{-1})_{mm} \ct^*_{mk}= \frac{1}{\sqrt{N}}\sum_{m=0}^{M}
\tbinom{2s+m-1}{m}^{1/2}(1-z\bz)^{s}\bz^{m}\lambda^{-1/2}_m
e^{2\pi ikm/N}\nn\\ &=&
\frac{1}{N}\left(\frac{1-z\bar{z}}{1-r^2}\right)^s\sum_{m=0}^M\left(\frac{\bar
z}{{re^{-2\pi ikm/N}}}\right)^{m}=\Xi_k(z),\eea
where we have used the expression of  $\langle z|s,m\rangle$ given
by eq. (\ref{upsilon}). $\blacksquare$

{\rem \label{lagrange} It is interesting to note that eq.
(\ref{reconstruccionover}) can be interpreted as a sinc-type
reconstruction formula, where the role of the sinc function  is
played by the function $\Xi_k(z)$, satisfying the ``orthogonality
relations" $\Xi_k(z_l)=P_{lk}$, where the operator $P=\ct \ct_l^+$
is an orthogonal projector onto a $M$-dimensional subspace of
$\mathbb{C}^N$, the range of $\ct$. In the case of critical
sampling, $N=M+1$, the result $\Xi_k(z_l)=\delta_{lk}$ is
recovered (corresponding to an interpolation formula), but for the
strict oversampling case, $N>M+1$, a projector is obtained to
account for the fact that an arbitrary set of overcomplete data
$\Psi_k,\,k=0,\ldots,N-1$, can be incompatible with
$|\psi\rangle\in {\cal H}_s^M$. $\square$}

A reconstruction in terms of the Fourier coefficients can be
directly obtained by means of the (left) pseudoinverse of the
frame operator $\ct$:

{\cor \label{corolarioover} {\rm (Discrete Fourier Transform)} The
Fourier coefficients $a_m$ of the expansion $|\psi\rangle =
\sum_{m=0}^M a_m|s,m\rangle$  of any $\psi\in{\cal H}_s^M$  can be
determined in terms of the data $\Psi_k=\langle
z_k|\psi\rangle$ as
\be a_{m}=\frac{1}{\sqrt{N\lambda_m}} \sum_{k=0}^{N-1}e^{2\pi
ikm/N}\Psi_k \,,\,m=0,\ldots,M\,. \label{coeffourierover} \ee }

\ni \textbf{Proof.} Taking the scalar product with $\langle z_k|$
in the expression (\ref{FourExpan}) of $|\psi\rangle$, we arrive
at the over-determined system of equations
\be \sum_{m=0}^{M} \ct_{km}a_{m}=\Psi_k,\;\;\ct_{km}=\langle
z_k|s,m\rangle, \label{sistema} \ee
which can be solved by left multiplying it by the (left)
pseudoinverse of $\ct$,
$\ct_l^+=(\ct^*\ct)^{-1}\ct^*=\ca^{-1}\ct^*$. Using the
expressions of $\ca^{-1}={\rm
diag}(\lambda_0^{-1},\lambda_1^{-1},\ldots,\lambda_{M}^{-1})$,
given in Lemma \ref{lemmaover1}, and the matrix elements
$\ct_{kn}$, given by the formula (\ref{upsilon}), we arrive at the
desired result.$\blacksquare$

{\rem Note that the Fourier coefficients $a_m$ are obtained as a
(rectangular) discrete Fourier transform of the data $\Psi(z_k)$
up to a rescaling factor  $1/\sqrt{\lambda_m}$ which can be seen
as a filter by $\ca^{-1/2}$. The expression
(\ref{coeffourierover}) provides a discretization of (\ref{CFT}).
$\square$}

\subsection{Band-Unlimited Functions and Undersampling}
In the previous section we have seen that, using $N$ sampling
points, we can fully reconstruct band-limited functions $\psi\in
{\cal H}^M_s$ of band-limit up to $M=N-1$. When the reconstruction
of a band-unlimited function $|\psi\rangle=\sum_{n=0}^\infty
a_n|s,n\rangle$ from a finite number $N$ of samples is required,
we cannot use the results of the previous section since the
resolution operator $\ca$ is no longer invertible\footnote{While
the operator $\ct:\ch_s\rightarrow \mathbb{C}^N$ has the same
expression as in the previous section, the operator $\ca$ is an
infinite dimensional matrix given by:
$\ca_{mn}=\lambda_{j+pN}^{1/2}\lambda_{j'+qN}^{1/2}\delta_{jj'}$,
with $m=j+pN$ and $n=j'+qN$, that is, it is a matrix made of
$N\times N$ diagonal blocks.}. However, the overlapping kernel
operator $\cb$ turns out to be invertible, and a partial
reconstruction can be done following the guidelines of the end of
Sec. \ref{CSandFrames} (undersampling).

Contrary to the case of the sphere \cite{samplingsphere}, where
the Hilbert space of functions of spin $s$, $\ch_s$, is
finite-dimensional, here $\ch_s$ is infinite-dimensional, and
therefore in the partial reconstruction of an arbitrary function
$|\psi\rangle=\sum_{n=0}^\infty a_n|s,n\rangle$ a considerable
error will be committed unless further assumptions on the Fourier
coefficients $a_n$ are made. Since $|\psi\rangle$ is normalizable,
the Fourier coefficients should decrease to zero, thus even if
$|\psi\rangle$ is not bandlimited, if $a_n$ decrease to zero fast
enough, it will be ``approximately'' band limited if the norm of
$|\psi_M^\perp\rangle\equiv \sum_{n=M+1}^\infty a_n|s,n\rangle$ is
small compared to  the norm of $|\psi\rangle$, for an
appropriately chosen $M$. Let us formally state these ideas.

{\defn Let us define by
\be P_M=\sum_{m=0}^M |s,m\rangle \langle s,m|\ee
the projector onto the subspace ${\cal H}_s^M$ of bandlimited
functions of bandlimit $M$. We shall denote by
\be \epsilon_{M+1}^2(\psi)\equiv {\cal E}_\psi^2({\cal H}^M_s)=
\frac{\langle\psi|I-P_M|\psi\rangle}{\langle\psi|\psi\rangle}=
\frac{\sum_{n=M+1}^\infty|a_n|^2}{\sum_{n=0}^\infty|a_n|^2}\,,\label{quasi-bandlimited}\ee
the normalized squared distance [similar to (\ref{errorpsi})] from
a band-unlimited function
\be |\psi\rangle=\sum_{n=0}^\infty a_n|s,n\rangle \in {\cal
H}_s\label{FourExpan2}\ee
to its orthogonal projection
\be |\psi_M\rangle=P_M|\psi\rangle=\sum_{n=0}^M a_n|s,n\rangle
\label{FourExpan3}\ee
onto the subspace ${\cal H}_s^M$. In other words,
$\epsilon_{M+1}(\psi)$ is the sine of the angle between $\psi$ and
$\psi_M$.}

We hope that the (normalized) error committed when reconstructing
$\psi$ from $N$ of its samples $\Psi_k$ will be of the order of
$\epsilon_{N}(\psi)$, which will be small as long as the Fourier
coefficients $a_n$ decay fast enough. More precisely, if
$|a_n|\leq C/n^\alpha$ for some constant $C$, $\alpha> 1/2$ and
$n\geq N$, then
\be\|\psi\|^2\epsilon_N^2(\psi)=\sum_{n=N}^\infty|a_n|^2\leq
C^2\sum_{n=N}^\infty\frac{1}{n^{2\alpha}}\leq
C^2\int_{N-1}^\infty\frac{1}{x^{2\alpha}}dx\leq
\frac{C^2}{2\alpha-1}\frac{1}{(N-1)^{2\alpha-1}},\label{asympepsilonN}\ee
which says that $\epsilon_N^2(\psi)=O(\frac{1}{N^{2\alpha-1}})$.
This condition could be more formally stated by saying that $\psi$
belongs to a certain Sobolev space $\mathbb H^k$ with
$k<\alpha-1/2$. In the next theorem we provide a partial
reconstruction formula for band-unlimited functions and a bound
for the error committed.

 {\thm \label{maintheoremunder} Given a
band-unlimited function $\psi\in {\cal H}_s$, there exists a
partial reconstruction of $\psi$, in terms of the alias
\be \hat{\Psi}(z)=\sum_{k=0}^{N-1}
L_k(z)\Psi_k,\label{reconstruccionunder} \ee
from $N$ of its samples $\Psi_k$, taken at the sampling points
in (\ref{discreteQ}), up to an error (\ref{errorpsi})
\be {\cal E}_\psi^2(r,N)\equiv {\cal E}_\psi^2({\cal H}^{\cal
S}_s) \leq
\frac{\nu_0(r,N)}{1+\nu_0(r,N)}+2\epsilon_{N}(\psi)\sqrt{1-\epsilon_{N}^2(\psi)}+
\epsilon_{N}^2(\psi)\frac{2+\nu_0(r,N)}{1+\nu_0(r,N)},
\label{errorpsi2}\ee
with $\nu_0(r,N)$ given by the formula (\ref{varepsilon}). The Lagrange-like interpolating functions (\ref{interpolating})
now adopt the following form:
\be
L_k(z)=\frac{1}{N}\left(\frac{1-z\bar{z}}{1-z_k\bar{z}_k}\right)^s\sum_{j=0}^{N-1}\hat{\lambda}_j^{-1}\sum_{q=0}^\infty
\lambda_{j+qN}\,(\overline{z z_k^{-1}})^{j+qN},\label{hatinterpolating}\ee
where
\be \hat{\lambda}_j=\sum_{q=0}^\infty
\lambda_{j+qN},\,j=0,\dots,N-1,\label{lambdagorro}\ee
are the eigenvalues of the discrete reproducing kernel operator
${\cal B}={\cal T}{\cal T}^*$ [defined in
(\ref{overlapping}) with matrix elements ${\cal
B}_{kl}=\langle z_k|z_l\rangle$] and $\lambda_n$ is given by
(\ref{lambdaover}), but now for $n=0,1,2,\ldots $\,. }

We shall see that the error (\ref{errorpsi2}) approaches zero when
$N\to \infty$. Before tackling the proof of this theorem, we shall
introduce some notation and prove some auxiliary results.

{\lem \label{lemmaunder1}  The pseudo-frame operator $\ct:{\cal
H}_s\to \mathbb C^N$ given by $\ct(\psi)=\{\langle
z_k|\psi\rangle, z_k\in {\cal Q}\}$ [remember the construction
after Eq. (\ref{pbiop2})] is such that the overlapping kernel
operator ${\cal B}=\ct\ct^*$ is an $N\times N$ Hermitian positive
definite invertible matrix, admitting the eigen-decomposition
${\cal B} ={\cal F}\hat{D}{\cal F}^*$, where $\hat{D} ={\rm
diag}(\hat\lambda_0,\ldots,\hat\lambda_{N-1})$ is a diagonal
matrix with $\hat\lambda_j$ given by (\ref{lambdagorro}) and
${\cal F}$ is the standard Fourier matrix.}

\ni \textbf{Proof.} Let us see that ${\cal B}$ is diagonalizable
and its eigenvalues $\hat\lambda_k$ are given by the expression
(\ref{lambdagorro}), which actually shows that all are strictly
positive and hence ${\cal B}$ is invertible. This can be done by
taking advantage of the circulant structure of ${\cal B}$ (see
e.g. Appendix B in \cite{samplingsphere}). Actually, using the
expression of the CS overlap (\ref{CSoverlap}), we have:
\be {\cal B}_{kl}=\langle
z_k|z_l\rangle=\left(\frac{1-r^2}{1-r^2e^{2\pi
i(l-k)/N}}\right)^{2s}\equiv{\cal C}_{l-k},\label{Bkl}\ee
where the circulant structure becomes apparent. The eigenvalues of
${\cal B}$ are easily computed by the formula:
\be \hat\lambda_k=\hat{D}_{kk}=({\cal F}^*{\cal B}{\cal
F})_{kk}=\frac{1}{N}\sum_{n,m=0}^{N-1}e^{i2\pi kn/N}{\cal
C}_{m-n}e^{-i2\pi mk/N}.\label{lambdak} \ee
If we expand the denominator of (\ref{Bkl}) in terms of binomial
coefficients,
\be {\cal C}_l=(1-r^2)^{2s}\sum_{q=0}^\infty
\binom{2s+q-1}{q}r^{2q}e^{2\pi i lq/N}\,,\ee
insert this in (\ref{lambdak}) and  use the general orthogonality relation for Rectangular
Fourier Matrices (\ref{exponencial}), we arrive at
(\ref{lambdagorro}). It is evident that $\hat{\lambda}_k>0,
\forall k=0,1,\ldots,N-1$, so that ${\cal B}$ is invertible.
$\blacksquare$

Following the general guidelines of Sec. 2, we now introduce the
projector $P_{\cal S}$:

{\lem \label{lemmaunder2} Under the conditions of the previous
Lemma, the set $\{|\tilde{z}_k\rangle =\sum_{l=0}^{N-1}
(\cb^{-1})_{lk}|z_l\rangle\,,k=0,\ldots,N-1\}$ constitutes a dual
pseudo-frame for ${\cal S}$, the operator $P_{\cal S}=\ct_r^+\ct$
is an orthogonal projector onto the subspace  ${\cal H}_s^{\cal
S}$, where $\ct_r^+=\ct^* \cb^{-1}$ is a (right) pseudoinverse for
$\ct$, and
\be
 \sum_{k=0}^{N-1} |\tilde{z}_k\rangle\langle z_k| =
\sum_{k=0}^{N-1} |z_k\rangle\langle \tilde{z}_k| =P_{\cal
S}\label{resolproy} \ee
provides a resolution of the projector $P_{\cal S}$, whose matrix
elements in the orthonormal base (\ref{ko}) of ${\cal H}_s$
exhibit a structure of diagonal $N\times N$ blocks:
\be P_{mn}(r,N)\equiv\langle s,m|P_{\cal S}|s,n\rangle
=(\lambda_m\lambda_n)^{1/2} \hat{\lambda}_{n \,{\rm mod}
N}^{-1}\delta_{(n-m) \,{\rm mod} N,0},
\;m,n=0,1,2,\dots,\label{projemn}\ee
with $\hat{\lambda}_n$ given by (\ref{lambdagorro}). }

\ni \textbf{Proof.} If we define $\ct_r^+=\ct^* \cb^{-1}$ it is
easy to check that $\ct \ct_r^{+}=I_N$ is the identity in
$\mathbb{C}^N$. In the same way, $P_{\cal S}=\ct_r^+\ct$ is a
projector since $P_{\cal
S}^2=\ct_r^+\ct\ct_r^+\ct=\ct_r^+\ct=P_{\cal S}$ and it is
orthogonal $P_{\cal S}^*=(\ct^* \cb^{-1}\ct)^* = \ct^* \cb^{-1}\ct
=P_{\cal S}$ since $\cb$ is self-adjoint. The resolution of the
projector is provided by Eq. (\ref{resolucionproyector}). Its
matrix elements can be calculated through:

\be P_{mn}(r,N)=\sum_{k,l=0}^{N-1}{\cal T}^*_{ml}({\cal
B}^{-1})_{lk}{\cal T}_{kn}.\label{Pmn}\ee
The inverse of ${\cal B}$ can be obtained through the
eigen-decomposition:
\be ({\cal B}^{-1})_{lk}=({\cal F}\hat{D}^{-1}{\cal
F}^*)_{lk}=\frac{1}{N}\sum_{j=0}^{N-1}\hat{\lambda}_j^{-1} e^{2\pi
ij(k-l)/N}.\label{inverseB}\ee
Inserting this last expression and ${\cal
T}_{kn}=\lambda_n^{1/2}{\cal F}_{kn}$ in (\ref{Pmn}) and using the
general orthogonality relation (\ref{exponencial}) for RFM, we
finally arrive at (\ref{projemn}). $\blacksquare$

The matrix elements (\ref{projemn}) will be useful when computing
the error function (\ref{errorpsi2}) for a general band-unlimited
function (\ref{FourExpan2}). At some point, we shall be interested
in their asymptotic behavior for large $N$ (large number of
samples). In order to give an explicit expression of this
asymptotic behavior of $P_{mn}(r,N)$, it will be useful to define
the following functions:
\be \nu_n(r,N)\equiv \frac{\hat{\lambda}_n-\lambda_n}{\lambda_n}=
\sum_{u=1}^{\infty}\frac{\binom{2s-1+n+uN}{n+uN}}{\binom{2s-1+n}{n}}r^{2uN},\;
n=0,\dots,N-1.\label{varepsilon}\ee
In terms of $\nu_n(r,N)$, the matrix elements (\ref{projemn})
adopt the following form:
\be
P_{mn}(r,N)=\frac{(\lambda_{j+pN}\lambda_{j+qN})^{1/2}}{\sum_{u=0}^\infty\lambda_{j+uN}}
=\frac{\binom{2s+j+pN-1}{j+pN}^{1/2}\binom{2s+j+qN-1}{j+qN}^{1/2}}
{\binom{2s+j-1}{j}}r^{(p+q)N}\frac{1}{1+\nu_j(r,N)},
\label{projemnepsilon} \ee
for $m=j+pN$ and $n=j+qN$, with $j=0,\dots,N-1$ and
$p,q=0,1,\dots$, and zero otherwise. In particular, for $n,m\leq
N-1$, the projector (\ref{projemnepsilon}) adopts the simple
diagonal form
\be P_{mn}(r,N)=\frac{1}{1+\nu_n(r,N)}\delta_{n,m},\;
n,m=0,\dots,N-1. \label{projemnepsilon1caja} \ee

Let us state and prove an interesting monotony property of
$\nu_n(r,N)$.

{\lem \label{lemmaunder4} The functions (\ref{varepsilon}) are
strictly decreasing sequences of $n$ for $r>0$, that is:
\be \nu_n(r,N)<\nu_m(r,N) \Leftrightarrow n>m,\;\;
n,m=0,\dots,N-1.\label{monotonyprop}\ee
}
\ni\textbf{Proof.}  The quotient of binomial coefficients
${\binom{2s-1+n+uN}{n+uN}}/{\binom{2s-1+n}{n}}$ in
(\ref{varepsilon}) is decreasing in $n$ for any $u\in \mathbb N$,
as can be checked by direct computation. Since this occurs for all
the coefficients of the power series in (\ref{varepsilon}), and
all of them are positive, the sequence $\nu_n(r,N)$ is decreasing
in $n$. $\blacksquare$

Now we are in conditions to prove our main theorem in this section.\\

\noindent\textbf{Proof of Theorem \ref{maintheoremunder}:}
According to (\ref{discrete-repker}), the pseudo-dual frame is
defined by
\be |\tilde{z}_k\rangle = \sum_{l=0}^{N-1} ({\cal
B}^{-1})_{lk}|z_l\rangle=\sum_{l=0}^{N-1}
\frac{1}{N}\sum_{j=0}^{N-1}\hat{\lambda}_j^{-1} e^{2\pi
ij(k-l)/N}|z_l\rangle.\nn\ee
Thus, the interpolating functions (\ref{interpolating}) read:
\be L_k(z)=\langle z|\tilde{z}_k\rangle=\sum_{l=0}^{N-1}
\frac{1}{N}\sum_{j=0}^{N-1}\hat{\lambda}_j^{-1} e^{2\pi
ij(k-l)/N}\langle z|z_l\rangle.\nn\ee
Noting that
\be \langle z|{z}_l\rangle=\sum_{n=0}^{\infty}\langle
z|s,n\rangle\langle s,n|{z}_l\rangle=\sum_{n=0}^{\infty}
\tbinom{2s+n-1}{n}^{1/2}(1-z\bz)^{s}\bz^{n}{\cal T}^*_{ln}, \ee
and using the orthogonality relation for Fourier matrices
(\ref{exponencial}), we arrive at (\ref{hatinterpolating}).

Now it remains to prove the bound (\ref{errorpsi2}) for the error.
Decomposing $|\psi\rangle$ in terms of $|\psi_{N-1}\rangle\equiv
P_{N-1}|\psi\rangle$ and $|\psi_{N-1}^\perp\rangle\equiv
(I-P_{N-1})|\psi\rangle$, we can write
\bea {\cal E}_\psi^2(r,N)\|\psi\|^2&=&\langle \psi|\psi\rangle  -
\langle \psi_{N-1}|P_{\cal S}|\psi_{N-1}\rangle - \langle \psi_{N-1}^\perp|P_{\cal S}|\psi_{N-1}^\perp\rangle\nn  \\
& & -2 {\rm Re} \langle \psi_{N-1}|P_{\cal
S}|\psi_{N-1}^\perp\rangle. \label{errordecomp}\eea
Let us start by bounding the term
\be \langle \psi_{N-1}|P_{\cal
S}|\psi_{N-1}\rangle=\sum_{n=0}^{N-1}|a_n|^2\langle n|P_{\cal
S}|n\rangle=\sum_{n=0}^{N-1}|a_n|^2\frac{1}{1+\nu_n(r,N)}\geq
\frac{1-\epsilon_{N}^2(\psi)}{1+\nu_0(r,N)}\|\psi\|^2, \ee
where we have used the expression (\ref{projemnepsilon1caja}),
Lemma \ref{lemmaunder4} in bounding $\frac{1}{1+\nu_n(r,N)}\geq
\frac{1}{1+\nu_0(r,N)}, \forall n=0,\dots,N-1$ and the definition
(\ref{quasi-bandlimited}). Next we shall bound the term
\bea  -2 {\rm Re} \langle \psi_{N-1}|P_{\cal
S}|\psi_{N-1}^\perp\rangle &\leq & 2|\langle \psi_{N-1}|P_{\cal
S}|\psi_{N-1}^\perp\rangle| \leq 2\|\psi_{N-1}\| \|P_{\cal
S}\psi_{N-1}^\perp\|\nn\\ &\leq & 2\|\psi_{N-1}\| \|P_{\cal
S}\|\|\psi_{N-1}^\perp\|=
2\sqrt{1-\epsilon_N^2(\psi)}\epsilon_N(\psi)\|\psi\|^2,
 \eea
where we have used that  $P_{\cal S}$ is an orthogonal projector,
therefore its spectral radius is $\rho(P_{\cal S})=\|P_{\cal
S}\|=1$, and the Cauchy-Schwarz inequality. Using the same
arguments, we can bound the remaining term as follows:
\be -\langle \psi_{N-1}^\perp|P_{\cal
S}|\psi_{N-1}^\perp\rangle\leq \langle \psi_{N-1}^\perp|P_{\cal
S}|\psi_{N-1}^\perp\rangle\leq \|\psi_{N-1}^\perp\| \|P_{\cal
S}\|\|\psi_{N-1}^\perp\|=\epsilon_N^2(\psi)\|\psi\|^2. \ee
Putting together all this information in (\ref{errordecomp}), we
arrive at the bound (\ref{errorpsi2}) for the error.$\blacksquare$

 It can
be easily seen that $\lim_{N\to\infty} \nu_0(r,N)=0, \forall r\in(0,1)$. As a consequence, the error (\ref{errorpsi2})
goes to zero as $N\to \infty$. To obtain the order of magnitude of this error, we shall firstly give
an asymptotic behavior of $\nu_0(r,N)$ for large $N$.

{\prop \label{lemmaunder3} The quantity $\nu_0(r,N)$ has the following asymptotic behavior (as a
function of $N$):
\be
\nu_0(r,N)=\binom{2s-1+N}{N}r^{2N}+O(N^{2s-1}r^{4N}),\label{varepsilon0asymp}
\ee
as long as
\be r<
r_0(s,N)=\left(\frac{\binom{2s-1+N}{N}}{\binom{2s-1+2N}{2N}}\right)^{\frac{1}{2N}}=
1-\frac{(2s-1)\ln 2}{2N}+O(1/N^2).\label{roptimo}\ee
}

\ni \textbf{Proof.} Let us see that the first addend ($u=1$) of
\be \nu_0(r,N)=
\sum_{u=1}^{\infty}\binom{2s-1+uN}{uN}r^{2uN}\label{varepsilon0}\ee
is dominant when $r<r_0(s,N)$. Indeed, the quotient between two
consecutive terms of the series (\ref{varepsilon0}) is
\be
r^{2N}\frac{\binom{2s-1+(u+1)N}{(u+1)N}}{\binom{2s-1+uN}{uN}}<r^{2N}
\frac{\binom{2s-1+2N}{2N}}{\binom{2s-1+N}{N}},\ee
where we have used that the quotient of binomials is decreasing in
$u\in\mathbb N$. If we impose the terms of the series
(\ref{varepsilon0}) to be monotonically decreasing for any
$u\in\mathbb N$, i.e.
\be r^{2N} \frac{\binom{2s-1+2N}{2N}}{\binom{2s-1+N}{N}}<1,\ee
then we arrive at (\ref{roptimo}). Thus, the first addend, $u=1$,
of (\ref{varepsilon0}) is the leading term. Using the Stirling
formula, the asymptotic behavior of the binomial
$\tbinom{2s-1+2N}{2N}$ of the second term, $u=2$, gives the
announced result (\ref{varepsilon0asymp}).  $\blacksquare$

Using the asymptotic expression (\ref{varepsilon0asymp}), the quadratic error (\ref{errorpsi2}) approaches zero
when $N\to\infty$, with the asymptotic behavior
\be {\cal E}_\psi^2(r,N)\leq
2\epsilon_{N}(\psi)\sqrt{1-\epsilon_{N}^2(\psi)}+
2\epsilon_{N}^2(\psi)+O(N^{2s-1}r^{2N}) \label{errorpsiasymp}\ee
for $r<r_0$. Thus, the reconstruction of $\psi$ by $\hat\psi$ is
exact in this limit.

{\cor {\rm (Discrete Fourier Transform)} The Fourier coefficients
$a_n$ of the expansion (\ref{FourExpan2}) can be approximated by
the discrete Fourier transform on the hyperboloid:
\be \hat{a}_n=\frac{\lambda_n^{1/2}}{\hat{\lambda}_{n {\rm mod}
N}} \frac{1}{\sqrt{N}}\sum_{k=0}^{N-1}e^{2\pi i
nk/N}\Psi_k,\label{dftunder}\ee
}

\ni \textbf{Proof.} The Fourier coefficients of the alias
(\ref{reconstruccionunder}) are given by:
 \be
 \hat{a}_n = \langle s,n|\hat{\psi}\rangle= \langle s,n|P_{\cal S}|\psi\rangle= \sum_{k=0}^{N-1}\langle s,n|\tilde
z_k\rangle\Psi_k=\sum_{k,l=0}^{N-1}\langle s,n|
z_l\rangle(\cb^{-1})_{kl}\Psi_k \,.\ee
Using that ${\cal T}_{ln}=\langle
z_l|s,n\rangle=\lambda_n^{1/2}{\cal F}_{ln}$, given in
(\ref{calt}), and the expression for the inverse of ${\cal B}$
given in (\ref{inverseB}), we obtain the final result once the
orthogonality relation (\ref{exponencial}) for Fourier Matrices is
used. $\blacksquare$

{\rem The expression for the Fourier coefficients $\hat{a}_n$
entails a kind of ``periodization" of the original $a_n$. Indeed,
putting
\[\Psi_k=\langle z_k|\psi\rangle=\sum_{m=0}^\infty \langle
z_k|m\rangle\langle m|\psi\rangle=\sum_{m=0}^\infty {\cal
T}_{km}a_m=\sum_{m=0}^\infty \lambda_m^{1/2}{\cal F}_{km}a_m\]
and using the orthogonality relations (\ref{exponencial}) we can
write (\ref{dftunder}) as:
\bea\hat{a}_n&=&\frac{\lambda_n^{1/2}}{\hat{\lambda}_{n {\rm mod}
N}} \sum_{k=0}^{N-1}{\cal F}_{nk}^*\sum_{m=0}^\infty \lambda_m^{1/2}{\cal F}_{km}a_m\nn\\
&=&\frac{\lambda_n^{1/2}}{\hat{\lambda}_{j}} \sum_{q=0}^\infty
\lambda^{1/2}_{j+qN}a_{j+qN},\;j=n\, {\rm mod} \,N,\eea
which implies
\be
\lambda_n^{-1/2}\hat{a}_n=\lambda_{n+pN}^{-1/2}\hat{a}_{n+pN}\Rightarrow
\hat{a}_{n+pN}=\sqrt{\frac{\lambda_{n+pN}}{\lambda_n}}\, \hat{a}_n\,,\,\forall
p\in\mathbb{N}\,. \ee
This is the ``hyperbolic'' counterpart of the typical
\emph{aliasing effect} for band-unlimited signals on the real
line.$\square$}

We could think that, for the case $\epsilon_{N}(\psi)=0$, we
should recover the results of Section \ref{oversampling}, but we
shall see that this is not the case. Before, a process of
truncation and filtering of $|\hat{\psi}\rangle$ in
(\ref{reconstruccionunder}) is necessary to recover the
reconstruction formula (\ref{reconstruccionover}) for strict
bandlimited functions (\ref{FourExpan}). Indeed, if $M=N-1$, the
truncation operation
\be |\hat{\psi}_M\rangle\equiv
P_M|\hat{\psi}\rangle=\sum_{m=0}^M\hat{a}_n|s,n\rangle\ee
followed by a rescaling (a filter) of the Fourier coefficients
\be |\hat{\psi}_M^R\rangle\equiv
R|\hat{\psi}_M\rangle=\sum_{m=0}^M\frac{\hat{\lambda}_n}{\lambda_n}\hat{a}_n|s,n\rangle\ee
renders the reconstruction formula for $\hat{\Psi}_M^R(z)=\langle
z|RP_M|\hat{\psi}\rangle$ to the expression
(\ref{reconstruccionover}). For band-unlimited functions, the new
bound for the squared normalized error turns out to be
\be \frac{\|\psi-\hat\psi^R_M\|^2}{\|\psi\|^2}\leq
\epsilon_N^2(\psi)+ \frac{\langle\psi_{M}^\perp|P_{\cal
S}P_MR^2P_MP_{\cal S}|\psi_{M}^\perp\rangle}{\|\psi\|^2}\leq
\epsilon_N^2(\psi)+\epsilon_N^2(\psi)(1+\nu_0(r,N))^2,\label{errorpsi3}\ee
where we have followed the same steps as in the proof of Theorem
\ref{maintheoremunder}, used that the spectral radius
$\rho(R)=\|R\|=1+\nu_0(r,N)$ and that $RP_MP_{\cal S}P_M=P_M$.
Contrary to (\ref{errorpsi2}), the new bound (\ref{errorpsi3}) is
proportional to $\epsilon_N^2(\psi)$. If we, moreover, assume a
behavior for $a_n$ as in (\ref{asympepsilonN}), then we have that
the error (\ref{errorpsi3}) is of order $O(1/N^{2\alpha-1})$.

Let us comment on an alternative approach to the sampling of
band-unlimited functions $\psi$ for small $\epsilon_{M+1}(\psi)$,
which will turn out to be more convenient in a certain limit.
Actually, for $\epsilon_{M+1}(\psi)<<1$ we have
\be
\|\psi-P_M\psi\|^2=\epsilon_{M+1}^2(\psi)\|\psi\|^2<<\|\psi\|^2,\ee
so that, the reconstruction formula (\ref{reconstruccionover}) for
$\psi_M=P_M\psi$ would give a good approximation of $\psi$,
similarly to the approach followed in \cite{Pesenson2-1}, Section
4. The problem is that, in general, the original data
$\Psi_k=\langle z_k|\psi\rangle$ for $\psi$ and the (unknown)
``truncated'' data $\Psi_{M,k}=\langle z_k|P_M|\psi\rangle$ for
$\psi_M$ are different unless $\langle z_k|P_M=\langle z_k|,
\forall k=0,\dots,N-1$, which is equivalent to $\langle
z_k|P_M|z_k\rangle=1, \forall k=0,\dots,N-1$. The following
proposition studies the conditions under which such requirement is
approximately satisfied.

{\prop For large symplin $s\to\infty$ and large band limit $M\to\infty$,
the diagonal matrix elements of $P_M$ in
 ${\cal H}_s^{\cal S}$ have the following asymptotic behavior:
\be P_M^s(p)\equiv \langle
z_k|P_M|z_k\rangle=\Theta(p_c-p)+O(\frac{1}{2s-1+M})\,,
\label{Heaviside} \ee
where $p\equiv r^2$ with $0\leq p<1$, $\Theta$ is the Heaviside
function,
\be
\Theta(x)=\left\{\begin{array}{lll} 0 & {\rm if} & x<0\\ 1 & {\rm
if} & x\geq 0\end{array}\right.
\ee
and
\be p_c=(1+\frac{2s-1}{M})^{-1}\label{radiocritico}\ee
denotes a critical squared radius. For $M>>2s$ we have
$p_c\lesssim 1$.

}

 \ni \textbf{Proof.} Using the expression (\ref{calt}) we have
\be P_M^s(p)\equiv \langle
z_k|P_M|z_k\rangle=\sum_{m=0}^M\ct_{km}\ct^*_{mk}=(1-p)^{2s}
\sum_{m=0}^M\binom{2s+m-1}{m}p^{m}.\ee
We can compute
\be  \frac{d P_M^s(p)}{d
p}=-(2s+M)\binom{2s+M-1}{M}(1-p)^{2s-1}p^M.\ee
We identify here the Binomial distribution $B(2s-1+M,p)$ (up to a
factor $2s+M$), which has a maximum (as a function of $p$) for
$p_c=1/(1+\frac{2s-1}{M})$. Using the Central Limit Theorem for
$2s-1+M\to\infty$ and the representation of the Dirac delta
function as the limit of a normal distribution, we identify
(\ref{Heaviside}) as a Heaviside-type (step) function,  concluding
the desired result.$\blacksquare$

\begin{figure}[h]
\begin{center}
\includegraphics[width=9cm,keepaspectratio]{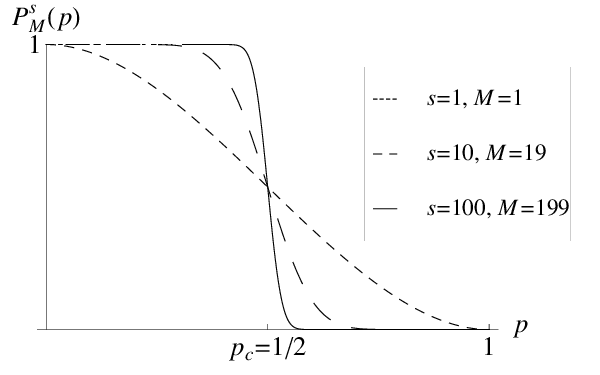}
\end{center}
\caption{\label{Droplet} $P_M^s(p)$ as a function of $p$ for
different values of $s$ and $M$, such that $p_c=\dfrac{1}{2}$.}

\end{figure}

Figure \ref{Droplet} shows a plot of $P_M^s(p)$ as a function of
$p$ for different values of $s$ and $M$ such that $p_c=1/2$, that
is, $M=2s-1$. It is clear how $P_M^s(p)$ approaches the step
function as $M$ and $s$ grow.

{\rem The matrix elements of $P_M$ in ${\cal H}_s^{\cal S}$ have a
circulant matrix structure. In fact, they can be seen as a Fourier
transform of the coefficients $\lambda_n$. They have the
expression $\langle z_k|P_M|z_l\rangle ={\cal C}_{k-l}(p)$, where
\be {\cal C}_l(p) = \frac{1}{N}\sum_{m=0}^M \lambda_m e^{-2\pi
i\frac{m l}{N}}\,. \ee
Note that ${\cal C}_{N-l}={\cal C}_l^*$, therefore the only
independents elements are ${\cal C}_l\,,l=0,\ldots,\frac{N}{2}$.
In the limit where $s$ and $M$ grow, $|{\cal C}_l(p)|$ rapidly
decreases to zero when $l$ approaches $\frac{N}{2}$, as can be
seen in Figure \ref{Droplet2}.

\begin{figure}[h]
\begin{center}
\includegraphics[width=8cm,keepaspectratio]{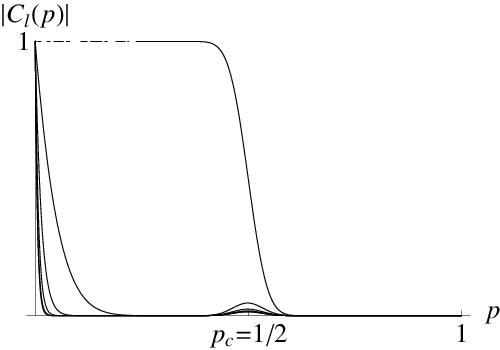}
\end{center}
\caption{\label{Droplet2} $|{\cal C}_l(p)|$  for $M=99$, $s=50$,
$N=100$ and different values of $l=0,10,20,30,40,50$.}

\end{figure}

\section{Conclusions and Outlook\label{comments}}

We have proved sampling theorems and provided DFT for holomorphic
functions on $\mathbb D_1$ carrying a unitary irreducible representation
of $SU(1,1)$ of symplin (Bargmann index) $s$. To accomplish our objective,
we used the machinery of Coherent States and discrete frames, and benefit
from the theory of Circulant Matrices and Rectangular Fourier Matrices to
explicitly invert resolution and reproducing kernel operators. We also
paved the way for more general coset spaces $Q=G/H$ and their
discretizations.

Heisenberg-Weyl (and Newton-Hooke) CS could be seen as a zero
curvature limit (and large $s$) of $SU(2)$ (positive curvature)
and $SU(1,1)$ (negative curvature) CS, a unified treatment of
sampling for the three cases being possible. This is left for
future work \cite{unified}.

\section*{Acknowledgements}
Work partially supported by the Fundación Séneca, Spanish MICINN
and Junta de Andalucía under projects [03100/PI/05, 08816/PI/08,
08814/PI/08], [FIS2005-05736-C03-01, FIS2008-06078-C03-01] and
FQM219, respectively. We thank the anonymous referees for useful
comments that have improved this paper and for bringing to our
attention some interesting references.


\footnotesize
\centerline{\rule{9pc}{.01in}}
\bigskip
\centerline{Departamento de Matem\'atica Aplicada y Estad\'\i
stica, Universidad Politécnica de Cartagena} \centerline{Paseo
Alfonso XIII 56, 30203 Cartagena, Spain} \centerline{e-mail:
Manuel.Calixto@upct.es}
\medskip
\centerline{Departamento de Matem\'atica Aplicada, Universidad de
Murcia, Facultad de Informática} \centerline{Campus de Espinardo,
30100 Murcia, Spain }

\centerline{e-mail: juguerre@um.es}
\medskip
\centerline{Departamento de Matem\'atica Aplicada y Estad\'\i
stica, Universidad Polit\'ecnica de Cartagena} \centerline{Paseo
Alfonso XIII 56, 30203 Cartagena, Spain} \centerline{e-mail:
JCarlos.Sanchez@upct.es}
\medskip

\end{document}